\documentclass[10pt,twocolumn,letterpaper]{article}

\usepackage{iccv}
\usepackage{times}
\usepackage{epsfig}
\usepackage{graphicx}
\usepackage{amsmath}
\usepackage{amssymb}
\usepackage{subcaption}

\usepackage[breaklinks=true,bookmarks=false]{hyperref}

\iccvfinalcopy 

\def\iccvPaperID{****} 

\ificcvfinal\fi

\begin{document}

\title{Visual Transformer with Statistical Test for COVID-19 Classification}

\author{$^1$Chih-Chung Hsu, $^2$Guan-Lin Chen, and $^3$Mei-Hsuan Wu\\
Institute of Data Science, National Cheng Kung University\\
No.1, University Rd., Tainan City, Taiwan ROC.\\
{\tt\small \{$^1$cchsu,$^3$re6091054\}@gs.ncku.edu.tw and $^2$alright1117@gmail.com} }

\maketitle
\ificcvfinal\thispagestyle{empty}\fi

\begin{abstract}
       With the massive damage in the world caused by Coronavirus Disease 2019 SARS-CoV-2 (COVID-19), many related research topics have been proposed in the past two years. The Chest Computed Tomography (CT) scans are the most valuable materials to diagnose the COVID-19 symptoms. However, most schemes for COVID-19 classification of Chest CT scan is based on a single-slice level, implying that the most critical CT slice should be selected from the original CT scan volume manually. We simultaneously propose 2-D and 3-D models to predict the COVID-19 of CT scan to tickle this issue. In our 2-D model, we introduce the Deep Wilcoxon signed-rank test (DWCC) to determine the importance of each slice of a CT scan to overcome the issue mentioned previously. Furthermore, a Convolutional CT scan-Aware Transformer (CCAT) is proposed to discover the context of the slices fully. The frame-level feature is extracted from each CT slice based on any backbone network and followed by feeding the features to our within-slice-Transformer (WST) to discover the context information in the pixel dimension. The proposed Between-Slice-Transformer (BST) is used to aggregate the extracted spatial-context features of every CT slice. A simple classifier is then used to judge whether the Spatio-temporal features are COVID-19 or non-COVID-19. The extensive experiments demonstrated that the proposed CCAT and DWCC significantly outperform the state-of-the-art methods.  
\end{abstract}

\section{Introduction}

With the rapid growth of the deep learning approach recently, the performance of many research fields has been boosted with deep learning. One essential application among them is medical image analysis based on deep learning. The chest Computed Tomography (CT) scan is an effective way to trace the symptoms of Coronavirus Disease 2019 SARS-CoV-2 (COVID-19). However, both the analysis and diagnosis of CT scan series require an experienced doctor or expert in the related field. Unfortunately, it is hard to meet this requirement in remote areas. An automatic computer-aid diagnosis system for CT scan for COVID-19 is thus highly desired. 


\begin{figure*}
     \centering
     \begin{subfigure}[t]{0.9\textwidth}
         \centering
         \includegraphics[width=0.9\linewidth]{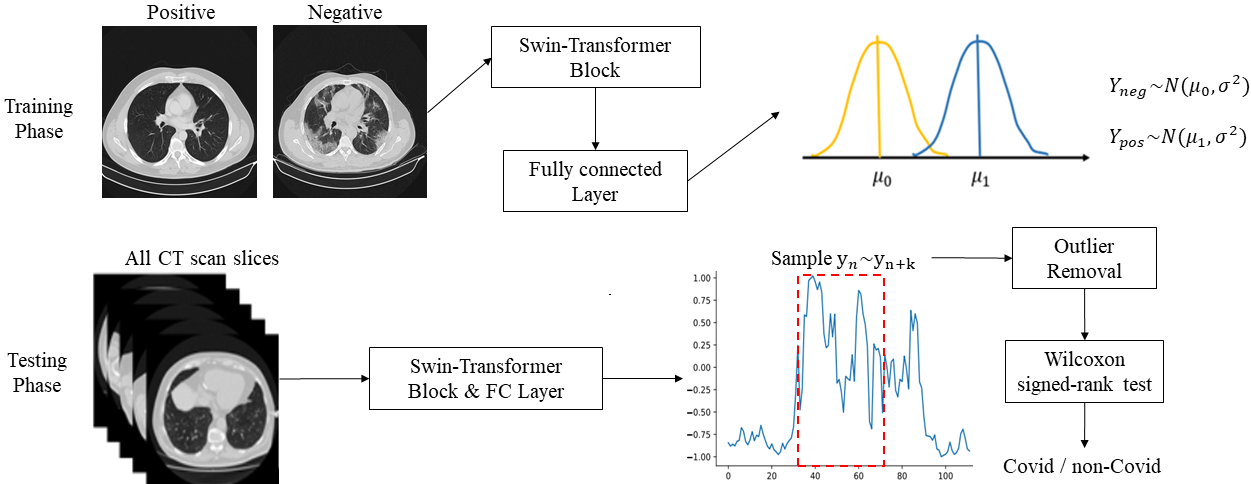}
         \caption{DWCC}
         \label{fig:dwcc}
     \end{subfigure}
     \begin{subfigure}[t]{0.9\textwidth}
         \centering
         \includegraphics[width=0.99\linewidth]{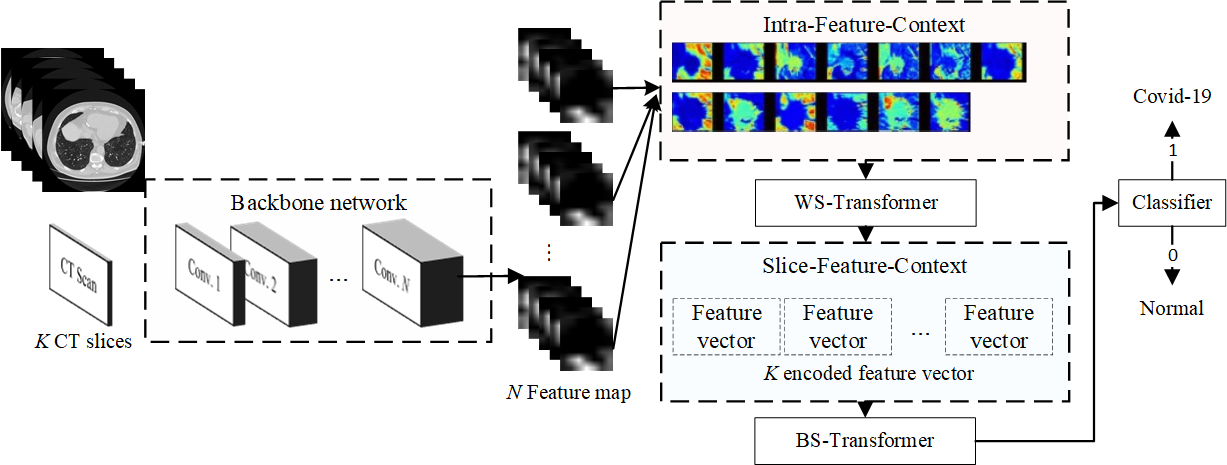}
         \caption{CCAT}
         \label{fig:ccat}
     \end{subfigure}
        \caption{Flowchart of the proposed methods (a) DWCC using statistical inference for deep learning and (b) CCAT for slice context and temporal relation mining modules.}
        \label{fig:flowchart}
\end{figure*}

The COVID-19 classification for CT scan can be modeled as a special case of the image/video recognition tasks. In the past decade, deep learning achieved state-of-the-art image recognition tasks compared to conventional machine learning and computer vision techniques. Similarly, deep learning-related schemes were widely adopted in the medical signal processing field \cite{ref4}. However, the CT scan series is usually treated as a three-dimensional (3-D) data volume, where the traditional convolutional neural network (CNN) can only perform the two-dimensional (2-D) convolutional operation on the image. 3-D convolution was then adopted to tickle these issues \cite{ct3d1}\cite{ct3d2}\cite{ct3d3}. However, the space and computational complexity of 3-D convolution is relatively higher than the 2-D, and therefore a high-end server is necessary.  Another critical issue is that the large-scale dataset is the power source of deep neural networks. The insufficient training samples lead to a fatal overfitting issue in training a deep neural network.

Compared to the 2-D information-only approach, the symptoms of COVID-19 might present at different depths (slice) for different patients, as suggested in \cite{prior1}. Fortunately, a large-scale 3D-shaped CT scan series, termed COV19-CT-DB, has been proposed in \cite{ref1}, including $5,000$ 3D CT scans with more than $1,000$ patients. Compared to the well-known traditional 2D CT dataset for COVID-19 classification \cite{ctdata}, the COV19-CT-DB contains 3D information providing more semantic features to help the diagnosis of COVID-19 symptoms. The conventional COVID-19 classification approach is usually based on an image-level scheme, where the input of their method is 2D image only \cite{ct0}\cite{ct1}\cite{ct2}\cite{ct3}\cite{ct4}\cite{ct5tip}\cite{ct6}. It is hard to adopt the 2-D information as the training set extends to the 3-D (i.e., CT scan series) without significant revision. The classification of COVID-19 based on a single slice can be treated as a conventional image recognition issue, as the suggestion in \cite{ct1}. In \cite{ct2}, self-supervision was proposed to improve classification performance for a small-scale COVID-19 CT scan dataset. Since the number of the training samples of CT scans is relatively tiny, supervised learning might be overfitting easily in practice. In \cite{ct4}, weakly supervised learning was introduced to tickle insufficient training samples and reduce the annotation requirements. However, the performance of those single slice-level models is restricted since the critical slice of a CT scan should be extracted by experienced experts. 

In \cite{ref2}\cite{ref3}, it can alleviate the catastrophic forgetting problem when another type of dataset related to the target disease has been used as the new training set. However, the 2-D and 3-D information are significantly different, implying that the performance might not be promising. The CT scan-level information provided in COV19-CT-DB makes it hard to perform these existing single slice-level schemes to 3-D CT scan-level directly without significant revision. Recently, several schemes were focused on CT scan-level (3-D information) directly. In \cite{ct3d1} adopted the multiple image-level CNNs to aggregate the predicted result of each CT image in a 3-D CT scan, while the 3D-CNN is directly adopted in \cite{ct3d2} to extract the cube-like feature from a whole 3-D CT scan. Both methods need considerable computational and space complexity to meet the model ensemble and 3-D convolution requirements. In \cite{ct3d4}, weakly supervised learning was adopted, as a suggestion in \cite{ct4}, to achieve lesion localization with classification annotation only based on an improved 3-D U-shape Network (U-Net). In \cite{ct3d3} and \cite{ref1}, the recurrent neural network (RNN) and long short-term memory (LSTM) network were proposed to integrate the cross-slice information to make the 3-D CT scan classification possible. However, it is well-known that the RNN and LSTM are hard to parallelize, leading to both training and inference time being hard to accelerate. Furthermore, 3-D convolution is significantly high computational complexity compared to 2-D convolution, leading to the fact that both training and inference costs are expensive.

Consider the computational complexity and effectiveness of CT scan-level classification of COVID-19; we respectively propose two schemes to resolve those issues based on the proposed maximum-likelihood estimation of Swin-Transformer \cite{swin} for slice-level classification and our Convolutional CT scan-Aware Transformer for CT scan-level classification.  

\section{The Proposed Method}
In this paper, two schemes are proposed based on the image-level (i.e., slice-level) and 3-D CT scan. In the first model, Swin-Transformer \cite{swin} is adopted as the backbone network for single-slice-level COVID-19 classification with Wilcoxon signed-rank test \cite{Rey2011}, termed as DWCC (Deep Wilcoxon signed-rank test for COVID-19 Classification). In the second model, we also explore the full 3-D information of the multiple slices of a CT scan series based on context feature learning, termed CCAT.

The flowchart of the first model is as shown in Fig.\ref{fig:flowchart} (\subref{fig:dwcc}). The positive and negative (i.e., COVID-19 and non-COVID-19) CT scan slices to different distribution by Swin-Transformer \cite{swin}, and determine the percentage in the middle of CT scan that has significant symptoms. Next, we apply outlier removal, and Wilcoxon signed-rank test \cite{Rey2011} for these generated samples. Wilcoxon signed-rank test can give meaning and explainable to the predicted results by statistical inference.

The entire flowchart of the proposed second approach is depicted in Fig.\ref{fig:flowchart} (\subref{fig:ccat}). The key components of the proposed CCAT are the Within-Slice-Transformer (WST) and Bewteen-Slice-Transformer (BST). The details of the proposed WST and BST will be introduced in the late subsection. In this WST, the training and testing CT scans $\mathbf{X}_t$ and $\mathbf{X}_v$ will be resized to a fixed size in the spatial domain as well as the $L_s$ slices will be sampled from the original CT scan series $\mathbf{X_{ct}}$. Afterward, a conventional CNN (ResNet-50 \cite{resnet} is used in this paper) is adopted to extract the feature maps $\mathbf{f} \in R^{c\times w_f\times h_f}$, where $w_f$ and $h_f$ indicate the width and height of $c$ feature maps. The global averaging pooling (GAP) is discarded to reserve the spatial information of $\mathbf{f}$. BST is adopted to explore how to aggregate the feature maps to context-encoded feature vector $f_{bst}$ based on self-attention mechanism. While the $L_s$ context-encoded features are aggregated $\mathbf{f}_t = [\textbf{f}_{bst}^0, \textbf{f}_{bst}^1,.., \textbf{f}_{bst}^{L_s}]$, the proposed WST is then used to mine the context features between features of slices $\mathbf{f}_{wst}$. Finally, a three-layer perceptron with LeakyReLU activation is designed as the classifier. 

\section{Experimental Result}
In this paper, the dataset used to evaluate the performance of the proposed approach is COV19-CT-DB \cite{ref1}. In COV19-CT-DB, the training and validation set are partitioned by \cite{ref1}, where the number of training and validation CT scans are $1,560$ and $374$, respectively. Since the annotation of the test set provided in \cite{ref1} is unavailable, validation set provided in \cite{ref1} is used to evaluate the performance of the proposed methods. 

In the training phase of our second model, the optimizer used in this paper is Adam \cite{adam}, the initial learning rate is $1e-4$ and the learning decay is step scheduler with step size $20$. The total epochs is $100$.

\subsection{Data pre-processing}
\subsubsection{DWCC}
\begin{figure}
		\centering
		\includegraphics[width=0.46\textwidth]{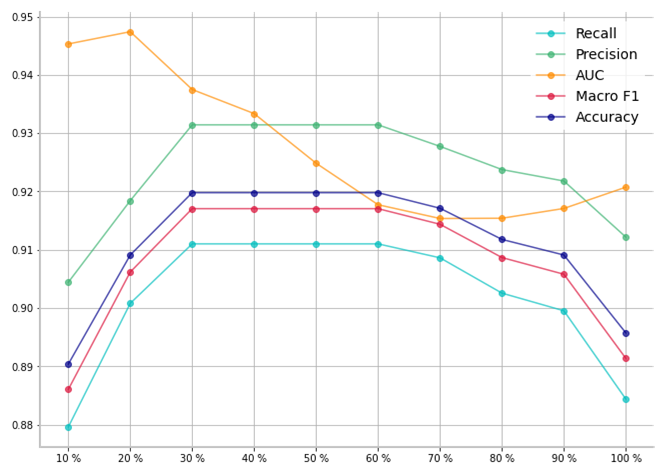}
		\caption{Validation performance of the proposed DWCC with different sample sizes.}
		\label{fig:sample_size}	
	\end{figure}

Due to some slices of CT scan might be useless for recognizing the COVID-19 (e.g., top/bottom slices might not contain chest information), the slices selection of the CT scan in the training phase is essential, as well as in the evaluation phase. In the training phase,  $40\%$  slices in the center of the CT scan are sampled, then augmentation and normalization will be performed on these selected slices. To empirically determine the best fraction of the slices selection, we conduct a performance based on different sampling sizes, as shown in Fig. \ref{fig:sample_size}. As a result, $30\%\sim60\%$ sample sizes in the center of CT scan make the best performance and therefore is suggested in our experiments.

\subsubsection{CCAT}
Since the number of the slices in each CT scan is significantly different from each other, the number of the input slices should be fixed to meet the requirements of our CCAT. In this paper, the number of slice $L_s=16$, and the sampling interval $L_{freq}=2$. We randomly sample $16$ slices in a CT scan to be an input data of the proposed CCAT. Meanwhile, the common data augmentation schemes, including blurring, noise, random contrast and brightness, and optical distortion, are adopted in the training phase. Note that the random rotation and cropping are not performed on each slice separately since it will lead to unstable of the context of slices of a CT scan. Therefore, the random cropping and rotation are performed on the sampled 3-D volume instead of each slice. Each pixel value will be normalized to be ranged $[0,1]$.

\subsection{Performance evaluation}

The evaluation metrics used in this paper are accuracy, macro-precision ($P$), macro-recall ($R$), and macro F1-score (F1). The performance comparison between the proposed and other peer methods is conducted in Table \ref{tab:result1}. It is clear that the proposed DWCC and CCAT significantly outperform the baseline model \cite{ref1} and other state-of-the-art model \cite{ct4}. As we stated previously, the single-slice-level model is hard to capture the characteristics of the slices' context, implying that the performance might be restricted. Although the Wilcoxon signed-rank test is proposed in DWCC to explore the importance of slices'\ automatically, the context of slices is still hard to address in practice. In contrast, the proposed CCAT, the proposed BS-transformer, further explores the context between slices, making the classifier more reliable and stable. Moreover, the backbone network used in our CCAT is ResNet-50 \cite{resnet} only since our WS- and BS-transformer can effectively extract the context features from CT scan. To further improve the performance, a model ensemble based on majority voting policy is adopted to fuse the predicted results of DWCC and CCAT, as shown in the last row at Table \ref{tab:result1}. As a result, the proposed CCAT significantly boosts the performance of the COVID-19 classification task compared to other state-of-the-art methods. 

\begin{figure*}
		\centering
		\includegraphics[width=1.0\textwidth]{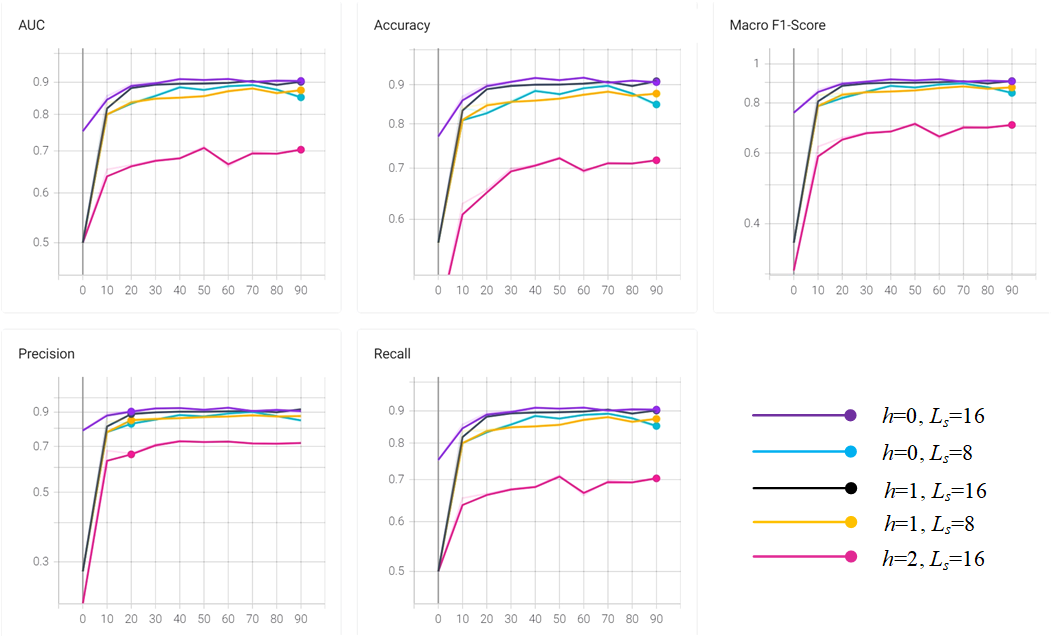}
		\caption{Validation performance curves of the proposed our CCAT model with different hyper-parameters in terms of (a) AUC, (b) Accuracy, (c) Macro F1-Score (evaluated by the official competition organization), (d) Macro Precision, and (e) Macro Recall.}
		\label{fig:curve}	
	\end{figure*}

To better understand the performance impact of the hyper-parameters of the proposed CCAT, a hyper-parameters tuning is illustrated in Fig. \ref{fig:curve}, where the $h$ is the number of the multi-head setting and the $L_s$ indicates the number of the slices. As a result, the length of retrieved slices of the CT scan and the number of the multi-head attention are suggested to $16$ and $0$ (i.e., gMLP \cite{gmlp}), respectively. All models presented a smooth and stable learning curve, implying that the proposed BS- and WS-Transformers can effectively learn the context features from the CT scan. 

\begin{table}
      \caption{Performance evaluation of validation set of the proposed methods and other peer methods in terms accuracy, macro-precision, macro-recall, macro-F1-score (results in {\color{blue}blue} indicates the implemented ourselves).}
      \centering
      \begin{tabular}{c|c|c|c|c}
        \hline
        Method & Acc. & $P$ &  $R$ & F1\\
        \hline
        Baseline \cite{ref1}        & {\color{blue}0.724} & {\color{blue}0.731} & {\color{blue}0.688} & 0.700 \\
        DenseNet201  \cite{ct4}     & 0.732 & 0.714 & 0.703 & 0.708 \\
        \hline
        Proposed DWCC                & 0.919 & 0.931 & 0.911 & 0.917 \\
        Proposed CCAT               & \textbf{0.933} & \textbf{0.935} & \textbf{0.929} & \textbf{0.932} \\
        \hline
        Model ensemble               & \textbf{0.941} & \textbf{0.947} & \textbf{0.935} & \textbf{0.939} \\
        \hline
    \end{tabular}\label{tab:result1}
    \end{table}
    
\section{Conclusion}
This paper has proposed two deep neural networks for two-dimensional (2-D) and three-dimensional (3-D) CT scan for COVID-19 classification tasks. First, the proposed DWCC (Deep Wilcoxon signed-rank test for COVID-19 Classification) adopts nonparametric statistics for deep learning, making the predicted result more stable and explainable, finding a series of slices with the most significant symptoms in CT scan. Second, the 3-D model has been proposed in this paper based on the pixel- and slice-level context mining, termed as CCAT (Convolutional CT scan-Aware Transformer), to further explore the intrinsic features in both temporal and spatial dimensions. Moreover, the proposed CCAT inherited the advantages of conventional CNNs and took the strengths of visual transformers. The visualization of the CT scan of the proposed CCAT models also verified that the critical insights of the symptoms caused by COVID-19 should be able to localize, as only CT scan-level annotation has been given. The extensive experiments have demonstrated that the proposed DWCC and CCAT significantly outperform the state-of-the-art methods for COVID-19 classification of CT scans.

{\small
\bibliographystyle{ieee_fullname}
\bibliography{egbib}
}

\end{document}